\documentclass[conference]{IEEEtran}
\IEEEoverridecommandlockouts
\usepackage[utf8]{inputenc}
\usepackage{amsmath,amssymb,amsfonts}
\usepackage{graphicx}
\usepackage{cite}
\usepackage{hyperref}
\usepackage{listings}
\usepackage{caption}
\usepackage{subcaption}
\usepackage{textcomp}
\usepackage{xcolor}
\usepackage{comment}
\usepackage[linesnumbered,ruled,vlined]{algorithm2e}
\def\BibTeX{{\rm B\kern-.05em{\sc i\kern-.025em b}\kern-.08em
    T\kern-.1667em\lower.7ex\hbox{E}\kern-.125emX}}
\SetKwBlock{Loop}{Loop}{EndLoop}

\makeatletter
\newcommand{\removelatexerror}{\let\@latex@error\@gobble}
\makeatother

\makeatletter
\newcommand{\linebreakand}{%
  \end{@IEEEauthorhalign}
  \hfill\mbox{}\par
  \mbox{}\hfill\begin{@IEEEauthorhalign}
}
\makeatother

\newcommand{\email}[1]{\href{mailto:#1}{#1}}

\title{%
LIFT: Byzantine Resilient  Hub-Sampling 
}

\author{\IEEEauthorblockN{Mohamed Amine LEGHERABA}
\IEEEauthorblockA{\textit{LIP6-NPA} \\
\textit{Sorbonne University}\\
Paris, France \\
\email{mohamed.legheraba@lip6.fr}}
\and
\IEEEauthorblockN{Nour RACHDI}
\IEEEauthorblockA{\textit{LIP6-NPA} \\
\textit{Sorbonne University}\\
Paris, France \\
\email{nour.rachdi@etu.sorbonne-universite.fr}}
\and
\IEEEauthorblockN{Maria POTOP-BUTUCARU}
\IEEEauthorblockA{\textit{LIP6-NPA} \\
\textit{Sorbonne University}\\
Paris, France \\
\email{maria.potop-butucaru@lip6.fr}}
\linebreakand
\IEEEauthorblockN{Sébastien TIXEUIL}
\IEEEauthorblockA{\textit{LIP6-NPA} \\
\textit{Sorbonne University}\\
\textit{IUF}\\
Paris, France \\
\email{sebastien.tixeuil@lip6.fr}}}

\begin{document}
\maketitle

\begin{abstract}

Recently, a novel peer sampling protocol, Elevator, was introduced to construct network topologies tailored for emerging decentralized applications such as federated learning and blockchain. Elevator builds hub-based topologies in a fully decentralized manner, randomly selecting hubs among participating nodes. These hubs, acting as central nodes connected to the entire network, can be leveraged to accelerate message dissemination. %
Simulation results have shown that Elevator converges rapidly (within 3–4 cycles) and exhibits robustness against crash failures and churn. However, its resilience to Byzantine adversaries has not been investigated.

In this work, we provide the first evaluation of Elevator under Byzantine adversaries and show that even a small fraction (2\%) of Byzantine nodes is sufficient to subvert the network.
As a result, we introduce LIFT, a new protocol that extends Elevator by employing a cryptographically secure pseudo-random number generator (PRNG) for hub selection, thereby mitigating Byzantine manipulation.  
In contrast, LIFT withstands adversarial infiltration and remains robust with up to 10\% Byzantine nodes. These results highlight the necessity of secure randomness in decentralized hub formation and position LIFT as a more reliable building block for Byzantine-resilient decentralized systems.
\end{abstract}


\section{Introduction}
Peer-to-peer (P2P) networks form the backbone of many modern distributed systems, allowing nodes to communicate and share resources without depending on centralized servers. Traditional peer sampling protocols like Cyclon~\cite{voulgaris2005cyclon}, PROOFS~\cite{proofs}, and Newscast~\cite{jelasity2007gossip} work by maintaining random connections between nodes, ensuring the network stays connected and spreads load evenly. While this approach works well, it creates flat network structures where all nodes have essentially the same role. New use cases in networking, such as blockchain~\cite{bitcoin2008bitcoin} or federated learning~\cite{mcmahan2017communication} necessitate the development of innovative peer sampling protocols that allows fast information broadcasting while remaining totally decentralized.

The Elevator protocol~\cite{legheraba2024emergent} takes a different approach by introducing hub-sampling—a method that naturally allows certain nodes to become hubs while keeping the network decentralized. Elevator combines two key ideas: preferential attachment (popular nodes get more connections) and random sampling (to maintain robustness). This creates networks with built-in hub structures that can spread information more efficiently and reduce the number of hops needed to reach any node. 

However, having nodes automatically become hubs raises new security concerns. More specifically, a fraction of malicious nodes that don't follow the protocol rules might try to exploit the hub selection process to gain more control over the network, that is, become hubs by a higher fraction. If successful, these attacks could undermine the network's security and the benefits that Elevator~\cite{legheraba2024emergent} is designed to provide.

Our contribution in this paper is twofold. First, we challenge Elevator's hub formation against Byzantine attacks. We study the impact of several attacks: a single Byzantine node that returns an empty cache or only its own address, several Byzantine nodes that are not coordinated, and finally several Byzantine nodes that are coordinated. Our simulation results show that Elevator is highly susceptible to Byzantine attacks when attackers are coordinated, and that as few as 2\% of Byzantine nodes are sufficient for all hubs to be Byzantine. Secondly,  we propose LIFT, an improvement to the Elevator protocol, resilient to Byzantine attacks while maintaining the same performances as Elevator protocol in terms of decentralization, and emergence of hubs. The LIFT defense mechanism activates automatically and instantly (all non-Byzantine nodes activate it at the same time). It allows a return to an acceptable threshold of Byzantine hubs in a single cycle.

The rest of this paper is structured as follows: Section~\ref{sec:context} explains the background on peer sampling and reviews existing work on Byzantine-resistant protocols. Section~\ref{sec:lift} describes our proposed counter attack against Byzantine failures. Section~\ref{sec:experiments} presents our experimental setup and demonstrates the effectiveness of our approach. and Section~\ref{sec:conclusion} concludes with future research directions.

\section{Context: Peer-sampling to Hub-Sampling}
\label{sec:context}

\subsection{Traditional Peer Sampling Approaches}

Peer sampling service is a fundamental building block in decentralized systems, particularly in large-scale peer-to-peer networks. It refers to the service used by each node in the network to maintains and periodically updates its local view containing a small, random subset of other nodes in the system. This service provides each participant with a continuously refreshed and approximately uniform sample of the global network, without requiring global knowledge or coordination. %
The output of the peer sampling layer is typically used to drive higher-level protocols, such as aggregation, clustering, or overlay construction.
A peer sampling service can be implemented either in a centralized or decentralized manner. In a centralized approach, a global coordinator maintains a complete view of the network and periodically provides each node with a partial view composed of randomly selected node identifiers. While this setup simplifies implementation, it introduces a single point of failure and a scalability bottleneck. In contrast, decentralized peer sampling protocols allow each node to construct and maintain its own local view by interacting with a small number of peers. The goal is to ensure that these local views collectively approximate a uniform random sampling of the global network. This decentralized design is preferable in fully distributed systems, as it eliminates reliance on any central authority and enhances fault tolerance. Gossip-based protocols are commonly used to achieve such decentralized and robust peer sampling. PROOFS~\cite{proofs} is a simple but effective protocol where nodes periodically exchange parts of their neighbor lists through shuffling. Nodes pick random neighbors to exchange information with, creating networks that behave like random graphs. Newscast~\cite{jelasity2007gossip} works similarly but adds the concept of "age" to node information—older connections are gradually replaced with fresher ones to maintain network quality. %

These traditional protocols all create flat networks where nodes have roughly equal roles and similar connection patterns. While this approach works well for basic connectivity and is robust against attacks targeting specific nodes, it misses opportunities for optimization in applications that could benefit from having some nodes serve as coordination points or  hubs.

\subsection{Byzantine-Resilient protocols}
Byzantine attack~\cite{byzantine} refers to arbitrary and potentially malicious behaviors by nodes in a distributed system. Unlike crash failures or omission faults, Byzantine nodes may deviate from the protocol in unpredictable ways, such as sending inconsistent messages to different peers, forging data, or coordinating with other malicious nodes to subvert the system. These attacks pose a significant threat to the robustness and correctness of distributed protocols, especially in decentralized environments where trust assumptions are minimal. In the context of peer sampling and hub selection protocols, Byzantine nodes can manipulate their local views to influence the overlay topology and to gain disproportionate visibility. Therefore, designing protocols that are resilient to such behaviors is essential for maintaining reliability, fairness, and convergence guarantees under adversarial conditions. When it comes to Byzantine fault-tolerant protocols, we can mention Phenix~\cite{wouhaybi2004phenix}, which allows the creation of power law networks. Phenix is resilient to  Byzantines that try to become hubs and cut off nodes from the network (targeted attacks). If a node detects that it's becoming disconnected from the network, it goes into “maintenance” mode and tries to create more connections with preferred nodes (the most connected nodes) to stay connected to the network. The Brahms protocol ~\cite{bortnikov2008brahms} uses a gossip peer sampling algorithm that is resistant to flood attacks, i.e. when a Byzantine node attempts to propagate its identifier to modify the composition of the local views of the network nodes, so as to be present in everyone's view. 
To do this, Brahms implements a sampling algorithm that manages to produce unbiased ID sampling from the history of received IDs, potentially biased by Byzantines, using min-wise independent permutations. \cite{anceaume2021byzantine} is in the same direction, but the algorithm that produces a uniform sample of IDs can adapt to a dynamic context. Basalt~\cite{basalt} is inspired by Brahms but uses a ranking function to select how its cache is updated. Secure Peer Sampling~\cite{jesi2010secure} is another proposition that is based on the gossip peer sampling protocol Newscast~\cite{jelasity2007gossip} but which adds an extra layer of security, with cryptography keys, a certificate authority and an algorithm to detects (potentially malicious) hubs.  SecureCyclon~\cite{antonov2023securecyclon} is based on Cyclon~\cite{voulgaris2005cyclon} and also add a layer of security that allow nodes to discover and blacklist malicious
nodes that violate the peer sampling protocol. AUPE~\cite{mukam2024aupe} leverage trusted nodes (like Intel SGX) to track the spread
of identifiers in the system.

These protocols are resilient to Byzantine attacks, but they only apply in cases where the network is a random graph. To our knowledge, there is no algorithm for resilience to Byzantine attacks that applies specifically to the Elevator case, i.e., where there are hubs in the network, and no studies have been conducted on Elevator's resilience to Byzantine attacks.
The Elevator protocol is an interesting protocol to study in that it is a completely decentralized protocol but at the same time offers the advantages of a centralized network by promoting certain nodes to hub status. Elevator is resilient to both outages and churn, however, the automatic hub promotion creates a potential security problem. Malicious nodes might try to exploit the preferential attachment process to artificially become hubs themselves, which could compromise the network's security and performance. Once they become hubs, Byzantines could take control of the application above the topology. For example, in the case of the HEAL protocol~\cite{legheraba2025heal}, which relies on Elevator for federated learning by using hubs as machine learning aggregators, if Byzantine nodes become hubs, they could hijack the learning process by injecting faulty machine learning models. In the absence of resilience to Byzantine attacks, the actual use of the protocol would be severely compromised.

\section{LIFT: Byzantine resilient hub sampling}
\label{sec:lift}

\subsection{Model of the system}
In the context of our study, we consider an overlay network of interconnected nodes modeled as a directed graph. Communication within this network is bidirectional, corresponding to an underlying undirected graph that represents the physical network. Each node in this network possesses a unique address (for example an IP address in the context of the Internet), serving as an abstract identifier of its identity. We assume that nodes cannot choose their identifiers (including malicious nodes) and that the identifier of each node is chosen randomly. Nodes maintain a local list called \emph{cache}, which contains addresses of other nodes, and represents their partial knowledge of the network's node set. The maximum size of this cache, denoted by parameter \emph{c}, is uniform across all nodes. At the network's inception, nodes are initially connected to a uniformly random subset of nodes, forming a random \emph{k}-out graph. We refer to the idea of \emph{cycles} of the protocol. During each cycle, every node initiates one execution of the peer sampling protocol, potentially updating its cache based on interactions with neighboring nodes (that is, with nodes in its cache at the beginning of the cycle).

\subsection{Elevator algorithm}
Network topologies that involve hubs are generally centralized, such as Gaia\cite{hsieh2017gaia}, where there are five servers that coordinate the learning work within the network. 
The Elevator protocol~\cite{legheraba2024emergent} takes a different approach by promoting some nodes to become hubs in the network, automatically and in a decentralized manner. Instead of keeping all nodes equal like traditional peer sampling protocols, Elevator creates networks where a fixed number of nodes (parameter $h$), chosen at random and in a decentralized manner, become connected to all the nodes in the network. Elevator combines two key ideas: preferential attachment (well-connected nodes get even more connections) and random connections (to keep the network robust). Each node keeps a cache of $c$ addresses of other nodes. During each round, nodes exchange their cache contents through gossip, and nodes with many connections gradually become hubs. This creates networks with very short paths between nodes (the diameter of the network is 2) and fast information spreading. This makes Elevator useful for applications like federated learning or blockchain networks that need to broadcast information quickly.

\begin{algorithm}[htbp]
\caption{Elevator Algorithm (active thread)}
\label{algo:algoLabel}
  \KwData{initial peer list: \emph{cache}}
  \KwData{cache size: \emph{c}}
  \KwData{number of hubs desired: \emph{h}}
  \KwData{initial backward list: \emph{$\mathit{backward\_peers}$} (empty)}
  \Loop{
    $\mathit{frequency\_{map}} \gets \{\}$\\ 
    $\mathit{preferred\_backward} \gets \{\}$\\ 
    \For{$\mathit{peer}\in \mathit{cache}$}{
        $\mathit{peer\_cache},\mathit{backward\_peer} \gets \mathit{send(CACHE\_{REQUEST},peer)}$\\
        $\mathit{frequency\_{map}} \gets \mathit{frequency\_{map}} \cup \mathit{peer\_cache}$\\
        $\mathit{preferred\_backward}.append(\mathit{backward\_peer})$\\
  	}
    $\mathit{preferred} \gets \mathit{frequency\_{map}.sortByFreq().select(number=h)}$\\    
    $\mathit{frequency\_{map}.remove(preferred)}$\\
    $\mathit{preferred\_backward.shuffle()}$\\
    $\mathit{cache} \gets \{\}$\\
    $\mathit{cache} \gets \mathit{preferred}[1..h] + \mathit{preferred\_backward}[1..c-h]$\\
    \While{$\mathit{cache.size()} < c$}{
        $\mathit{peer} \gets \mathit{frequency\_{map}.selectRandom()}$ \\
        $\mathit{cache.append(peer)}$\\
    }
  }
\end{algorithm} 

\begin{algorithm}[htbp]
\caption{Elevator Algorithm (background thread)}
\label{algo:algoLabelBackground}
  \Loop{
    $\mathit{request, peer} \gets \mathit{receive()}$\\
    \If{$\mathit{request} = \mathit{CACHE\_{REQUEST}}$}{
        $ \mathit{backward\_peers.add(peer)}$\\
        $ \mathit{backward\_peers.shuffle()}$\\
        $ \mathit{send(cache, backward\_peers[0], peer)}$\\
        }
}
\end{algorithm}

Here we review how the Elevator protocol works. The protocol executes the following actions at each run: Each node retrieves the neighbor's list of their neighbors (\emph{i.e.}, the successors at distance two). The node then builds an ordered list of the most frequent peers (the frequency map) and contacts the \emph{h} most frequent nodes (called \emph{preferred}). Each contacted node sends back to the contacting node a random address from its backward peer list, and adds the contacting node to its backward list. The cache of the contacting node is then reset as an empty array. Then the node selects the \emph{h} most frequent peers and \emph{c-h} random peers from the list of backward peers. If the cache is not full, the node adds random peers from the frequency map to the cache until the size of the cache is \emph{c} (see Algorithm~\ref{algo:algoLabel} and Algorithm~\ref{algo:algoLabelBackground} for detailed pseudocode of the algorithm).

\subsection{Byzantine attacks model}
Our attack model assumes that a certain percentage of nodes are Byzantine from the start. These malicious nodes try to break the Elevator protocol by sending false information during cache exchanges, manipulating the hub selection process. The goal of Byzantine attackers is to get selected as hub by the correct nodes (that genuinely execute the protocol). Obviously, if there is a fraction $p$ of Byzantine nodes overall, then it is trivial for the Byzantine nodes to obtain a fraction $p$ of the hubs (they should just behave as correct nodes). So, the Byzantine nodes strive to obtain a higher fraction of the hubs than their fraction of the nodes. 

\textbf{Attack Mechanism:} When legitimate nodes ask Byzantine nodes for their cache contents or backward peers information (i.e: the nodes who contacted them in the past), the Byzantine nodes respond with fake data designed to help malicious nodes become hubs. This attack works because Elevator relies on nodes honestly reporting their connectivity information. Apart from that, Byzantine nodes perform the protocol like other nodes. Byzantine nodes modify their behavior in order to achieve the goal of having a large proportion of hubs be Byzantine, but their objective is also to avoid detection. If their behavior deviates too much from that of a normal node, they could easily be detected and blacklisted.

We are studying several type of Byzantine nodes:
\begin{enumerate}
    \item Passive unique byzantine: A single Byzantine sending an empty cache
    \item Active unique byzantine: A single Byzantine who sends his modified cache with a reference to himself (to increase his probability of being chosen as a hub)    \item Non-coordinating byzantine nodes : Multiple byzantine nodes who sends their modified cache with a reference to themselves but the don't have references to other byzantine nodes
    \item Coordinated byzantine nodes : Each Byzantine node maintains a coordinated fake cache containing references to all other Byzantine participants in the network. When responding to legitimate cache requests, Byzantine nodes return sublists of this coordinated cache, effectively creating an artificial preference for Byzantine nodes in the sampling process.
\end{enumerate}

\begin{algorithm}[htbp]
\caption{Do-nothing attack}
\label{algo:DoNothingAttack}
  \Loop{
    $\mathit{request, peer} \gets \mathit{receive()}$\\
    \If{$\mathit{request} = \mathit{CACHE\_{REQUEST}}$}{
        $ \mathit{backward\_peers.add(peer)}$\\
        $ \mathit{backward\_peers.shuffle()}$\\
        $ \mathit{send([], None, peer)}$\\
        }
}
\end{algorithm}
\begin{algorithm}[htbp]
\caption{Non-coordinating attack}
\label{algo:NonCoordinatingAttack}
  \KwData{my address: \emph{my\_address}}
  \Loop{
    $\mathit{request, peer} \gets \mathit{receive()}$\\
    \If{$\mathit{request} = \mathit{CACHE\_{REQUEST}}$}{
        $ \mathit{backward\_peers.add(peer)}$\\
        $ \mathit{backward\_peers.shuffle()}$\\
        $ \mathit{modified\_cache} \gets cache.remove(random()).add(my\_address)$\\
        $ \mathit{send(modified\_cache, my\_address, peer)}$\\
        }
}
\end{algorithm}
\begin{algorithm}[htbp]
\caption{Coordinated attack}
\label{algo:CoordinatedAttack}
  \KwData{addresses of all byzantines nodes: \emph{all\_byzantines}}
  \Loop{
    $\mathit{request, peer} \gets \mathit{receive()}$\\
    \If{$\mathit{request} = \mathit{CACHE\_{REQUEST}}$}{
        $ \mathit{backward\_peers.add(peer)}$\\
        $ \mathit{backward\_peers.shuffle()}$\\
        $ \mathit{all\_byzantines.shuffle()}$\\
        $ \mathit{modified\_cache} \gets all\_byzantines[0:c]$\\
        $ \mathit{random\_backward} \gets all\_byzantine.random\_value()$\\
        $ \mathit{send(modified\_cache, random\_backward, peer)}$\\
        }
}
\end{algorithm}

In terms of pseudo-code, this amounts to replacing the Algorithm \ref{algo:algoLabelBackground} (the background Elevator process) with the following algorithms: Algorithm \ref{algo:DoNothingAttack} for the passive unique byzantine, Algorithm \ref{algo:NonCoordinatingAttack} for the active unique byantine attack and the multiple non coordinating byzantines, and Algorithm \ref{algo:CoordinatedAttack} for the multiple coordinated byzantines.

The Elevator protocol was not designed to be resilient to Byzantine attacks, and the protocol assumes that each node is honest and returns reliable information. Since in Elevator each node modifies its cache based on the cache of its neighbors, having one or more Byzantine nodes among its neighbors significantly changes the local behavior of the protocol (for a given node) and therefore the overall convergence towards the h hubs. It is therefore necessary to consider an alternative algorithm, based on Elevator but which takes into account the possibility of Byzantine attacks, while remaining decentralized.

\subsection{LIFT protocol}

To address Elevator's vulnerability to Byzantine attacks, we propose a deterministic hub redistribution mechanism that activates after the network has converged to its initial hub configuration. Our approach leverages the fact that node identifiers are assigned randomly and cannot be modified by Byzantine nodes. If Byzantine nodes are active, we hope that our new protocol will be more efficient than Elevator in terms of resilience, and if Byzantine nodes are not active, we hope that the protocol will have no impact on protocol performance and convergence towards hubs.

The counter-attack operates in two phases: an initial convergence phase using standard Elevator, followed by a deterministic hub redistribution phase.

\textbf{Phase 1 - Initial Convergence:} The network runs the standard Elevator protocol for a predetermined number of cycles (100 cycles in our implementation) to allow hub formation. We would like to point out that, according to simulation results\cite{legheraba2024emergent}, the Elevator protocol converges on average in 4 cycles, and therefore 100 cycles is more than enough time to ensure that we have reached convergence, which corresponds to having a network topology with $h$ nodes that are in everyone's cache, and the rest of the nodes' cache filled with random identifiers of other nodes. During this phase, Byzantine nodes may successfully infiltrate hub positions through coordinated attacks.

\textbf{Phase 2 - Hub Redistribution:} After convergence, all correct nodes simultaneously execute the following deterministic process (see Algorithm~\ref{algo:lift} for detailed pseudocode of the algorithm): \emph{i)} Each correct node first retrieves the list of the identifiers of the $h$ current hubs (which are potentially Byzantine). To obtain this list, each correct node simply needs to look at the first $h$ elements of its cache, since after Elevator convergence, the first $h$ elements of each correct node's cache correspond to the addresses of the $h$ hubs (a hub can have itself in its cache). \emph{ii)} Each node builds a seed by concatenating the $h$ identifiers of the hubs. The $h$ identifiers are already sorted, so the seed obtained is the same for every correct node. \emph{iii)} Each node initializes a pseudo-random number generator (PRNG) using the seed. The PRNG used is the default one in Java, and therefore a linear congruential pseudorandom number generator~\cite{knuth1997taocp3}, and is the same for everyone, as is the seed, so the numbers obtained at the PRNG output are the same for every correct node, effectively creating a shared random list of numbers. \emph{iv)} Each correct node uses the PRNG to generate $h$ new random values, corresponding to $h$ node identifiers in the network. If a value has already been obtained, the PRNG is used again until a new value is obtained. \emph{v)} Each correct node replaces the first $h$ identifiers (corresponding to $h$ potentially Byzantine hubs) in its cache with the $h$ identifiers obtained by the PRNG. It therefore removes its old connections to the old hubs (possibly invaded by Byzantine nodes) and replaces them with the new hubs (chosen uniformly at random).

\begin{algorithm}[htbp]
\caption{LIFT : Deterministic Hub Redistribution}
\label{algo:lift}
  \KwData{current hub list: $\mathit{H}$}
  \KwData{network size: $\mathit{N}$}
  \KwData{target hubs: $\mathit{h}$}
  
  $\mathit{hubIDs} \gets getSortedHubIDs(\mathit{H})$ \tcp*{Extract and sort hub node IDs}
  $\mathit{seed} \gets hashCode(\mathit{hubIDs})$ \tcp*{Generate deterministic seed}
  $\mathit{prng} \gets Random(\mathit{seed})$ \tcp*{Initialize PRNG with seed}
  
  $\mathit{selectedIDs} \gets \{\}$ \;
  $\mathit{newHubs} \gets \{\}$ \;
  
  \While{$|\mathit{selectedIDs}| < \mathit{h}$}{
    $\mathit{randomID} \gets \mathit{prng}.nextInt(\mathit{N})$ \tcp*{Random node ID in $[0, N-1]$}
    \If{$\mathit{randomID} \notin \mathit{selectedIDs}$}{
      $\mathit{targetNode} \gets network.get(\mathit{randomID})$\;
      \If{$\mathit{targetNode} \neq null \land \mathit{targetNode}.isUp()$}{
        $\mathit{selectedIDs} \gets \mathit{selectedIDs} \cup \{\mathit{randomID}\}$\;
        $\mathit{newHubs} \gets \mathit{newHubs} \cup \{\mathit{targetNode}\}$\;
      }
    }
  }
  
  $replaceCache(\mathit{newHubs}, \mathit{currentNode})$ \tcp*{Update cache with new hubs}
\end{algorithm}

Since all nodes use the same seed derived from the initial hub selection, they deterministically select identical new hub sets. Because node identifiers are randomly assigned and immutable, each node has equal probability $\frac{h}{N}$ of becoming a hub, regardless of Byzantine status.

Our implementation activates the counter-attack at cycle 100, allowing sufficient time for initial hub formation while preventing Byzantine nodes from establishing permanent control. The algorithm replaces the cache contents entirely: the first $h$ positions are filled with the deterministically selected new hubs, while remaining positions are populated with random non-hub nodes to maintain cache diversity.

The critical insight is that Byzantine nodes cannot manipulate their node identifiers, which are assigned during network initialization. Therefore, even if Byzantine nodes dominate the initial hub selection process, the subsequent deterministic redistribution treats all nodes equally based on their immutable identifiers.

\section{Experimental evaluation}
\label{sec:experiments}

\subsection{Experimental Framework}
We use PeerSim~\cite{peersim}, a simulator designed for P2P protocol analysis, to test Elevator's security. We modified the original Elevator implementation to add Byzantine behavior.

\textbf{Network Configuration:} All tests use networks of 1000 nodes with cache size $c = 20$ and target hub count $h = 10$. We use a K-out topology where nodes initially have K random identifiers of other nodes in their cache (K=c=20), which simulates how real P2P networks start up. The number B of Byzantine nodes is precisely specified in the configuration files for each experimental scenario. 
 
\textbf{Evaluation Metrics:} We measure Elevator's Byzantine resilience using two key metrics: (i) \textit{Hub formation rate}—how many hub positions are held by legitimate nodes versus attackers (Byzantine nodes), and (ii) \textit{Network topology stability}—whether hub formation still works properly under attack.

Each test runs for 1000 cycles to ensure the network stabilizes, and we average results over 100 independent simulations to account for randomness in network setup and protocol execution.

\subsection{Impact of byzantine attacks on Elevator}
We first evaluate the impact of Byzantine attacks on Elevator and then the effectiveness of the LIFT countermeasure protocol.

\begin{figure}[htbp]
\centering
\includegraphics[width=0.45\textwidth]{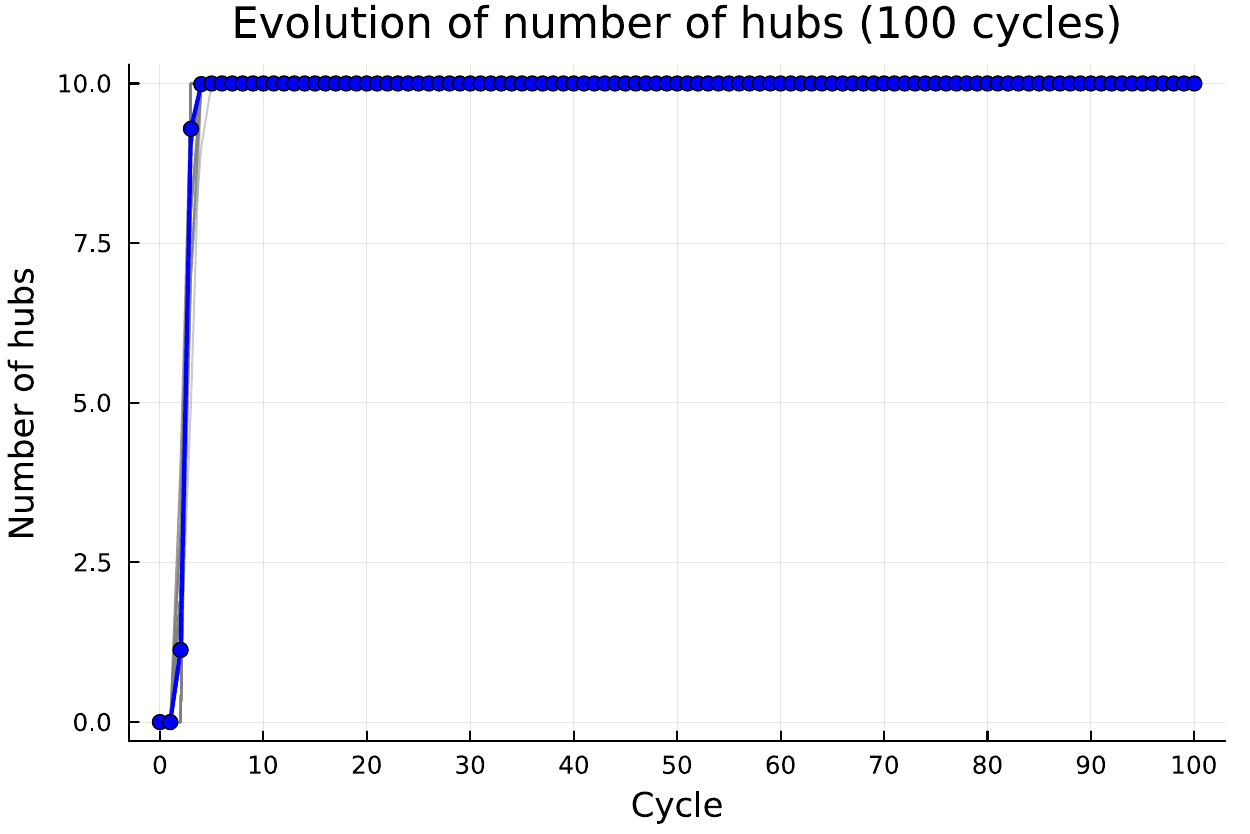}
\caption{Running of Elevator without attack. The convergence to the hub happens in less than 4 cycles.}
\label{fig:no_attack}
\end{figure}

As shown in Figure~\ref{fig:no_attack}, when the Elevator protocol is running normally (without byzantine nodes), we have the convergence to the 10 hubs very quickly, in less than 4 cycles on average.

\subsubsection{Individual byzantine node}
We analyze the impact of a single Byzantine node. In the network of 1,000 nodes, there is only one Byzantine node whose objective is to become a hub.

\begin{figure}[htbp]
\centering
\begin{subfigure}[b]{0.45\textwidth}
\includegraphics[width=\textwidth]{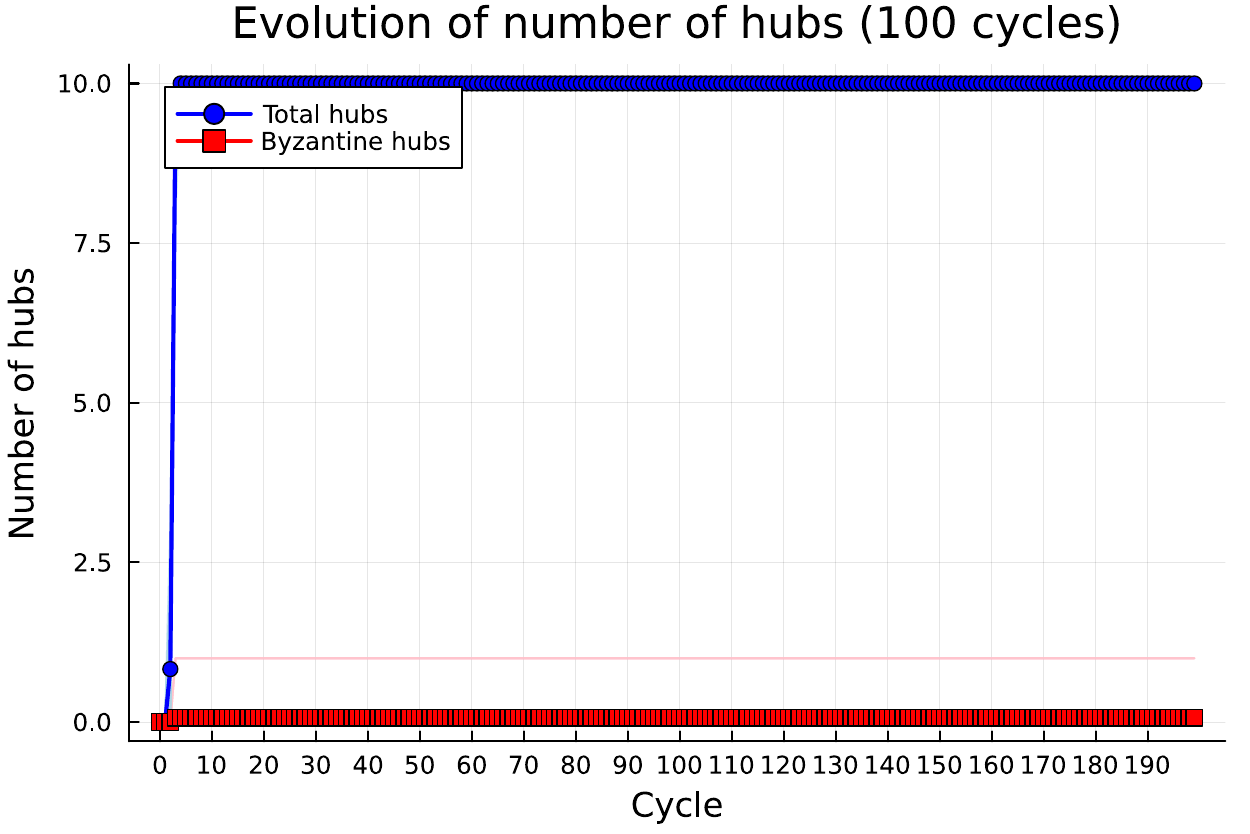}
\caption{Active Byzantine behavior}
\label{fig:single_byzantine_active}
\end{subfigure}
\hfill
\begin{subfigure}[b]{0.45\textwidth}
\includegraphics[width=\textwidth]{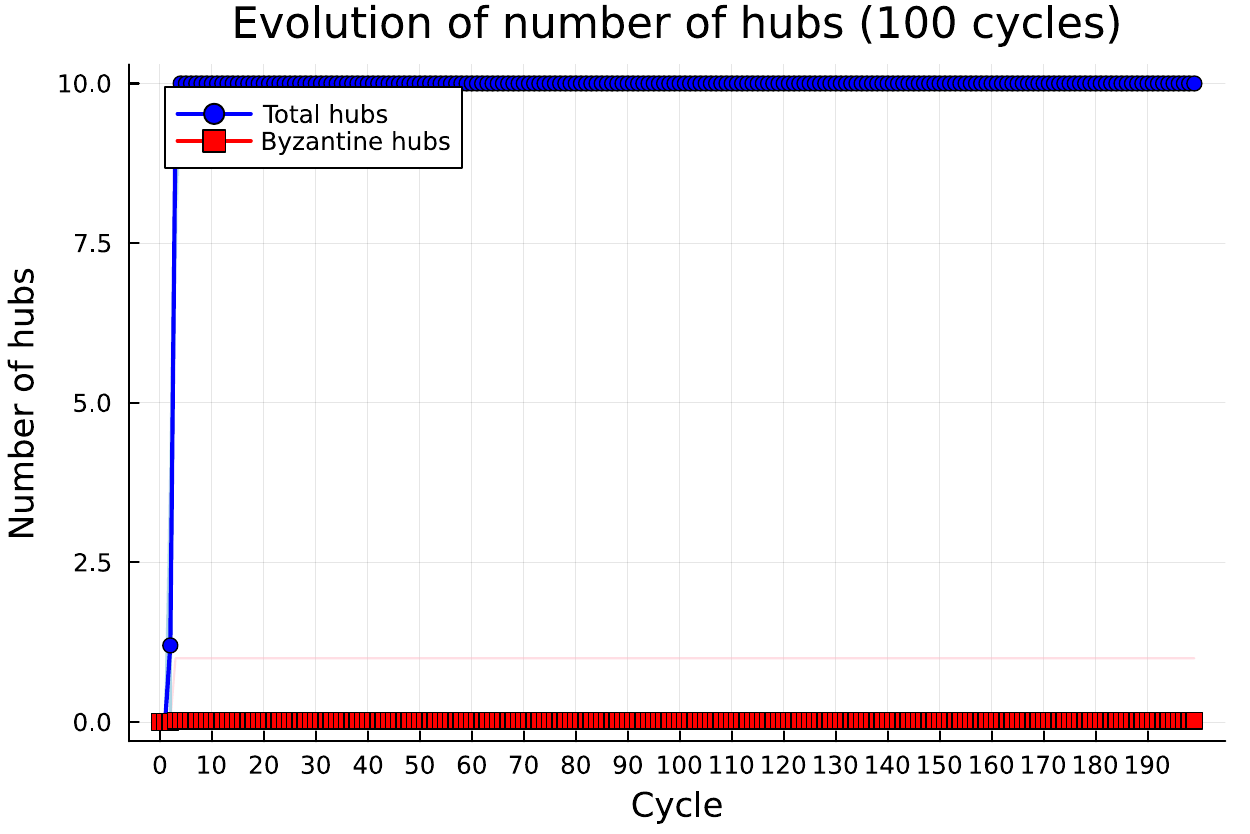}
\caption{Passive Byzantine behavior}
\label{fig:single_byzantine_passive}
\end{subfigure}
\caption{Impact of single Byzantine node with different attack strategies on hub formation in 1000-node networks.}
\label{fig:single_byzantine}
\end{figure}

Figure~\ref{fig:single_byzantine} shows that individual Byzantine nodes, regardless of their specific attack strategy, cannot significantly impact hub formation in networks of 1000 nodes. For the passive byzantine, the malicious node has been a hub 2 times out of 100 and the number of hubs is always 10. For the active byzantine, the malicious node has been a hub 7 times out of 100 and the number of hubs is also always 10. This result provides confidence in Elevator's resistance against low-level attacks and individual bad actors.

\subsubsection{Non-coordinating byzantine nodes}
We now look at the impact of an attack by several uncoordinated Byzantine nodes. The Byzantine nodes are placed randomly in the network.

\begin{figure}[htbp]
\centering
\includegraphics[width=0.45\textwidth]{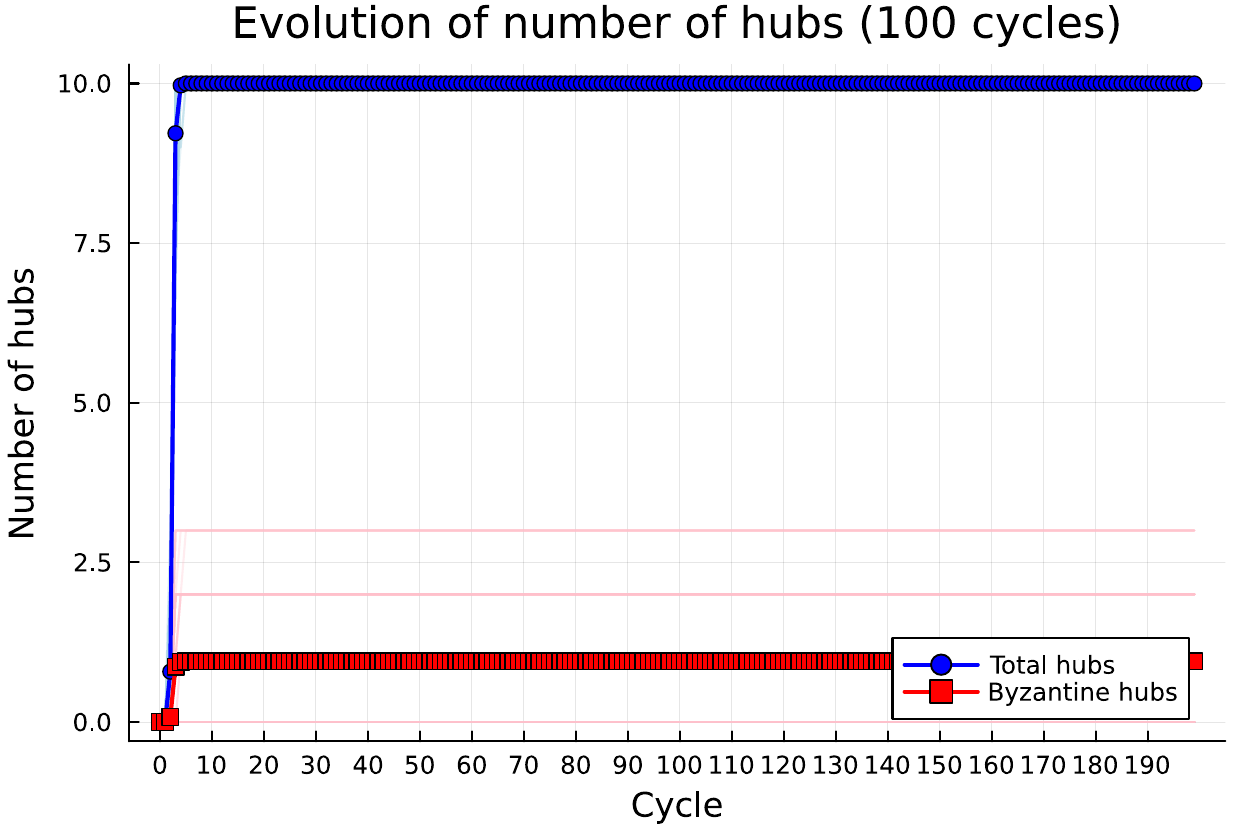}
\caption{Independent Byzantine attack at 5\% participation showing minimal hub infiltration due to lack of coordination among attackers.}
\label{fig:independent_byzantine}
\end{figure}

Our results, presented in Figure~\ref{fig:independent_byzantine}, demonstrate that independent Byzantine nodes have minimal impact on hub formation. On average, we have 0.95 out of 10 hubs that are Byzantine. We therefore go from 5\% Byzantines among nodes to 9.5\% among hubs, which is significant but still acceptable.
This finding highlights the critical importance of coordination in Byzantine attacks against Elevator's hub selection mechanism.

\subsubsection{Coordinated byzantine nodes}
We will now examine the impact of an attack by several coordinated Byzantine nodes (each Byzantine node knows the list of all other Byzantine nodes and shares this list when asked for its cache). Here too, the Byzantine nodes are placed randomly in the network.

\begin{figure}[htbp]
\centering
\includegraphics[width=0.45\textwidth]{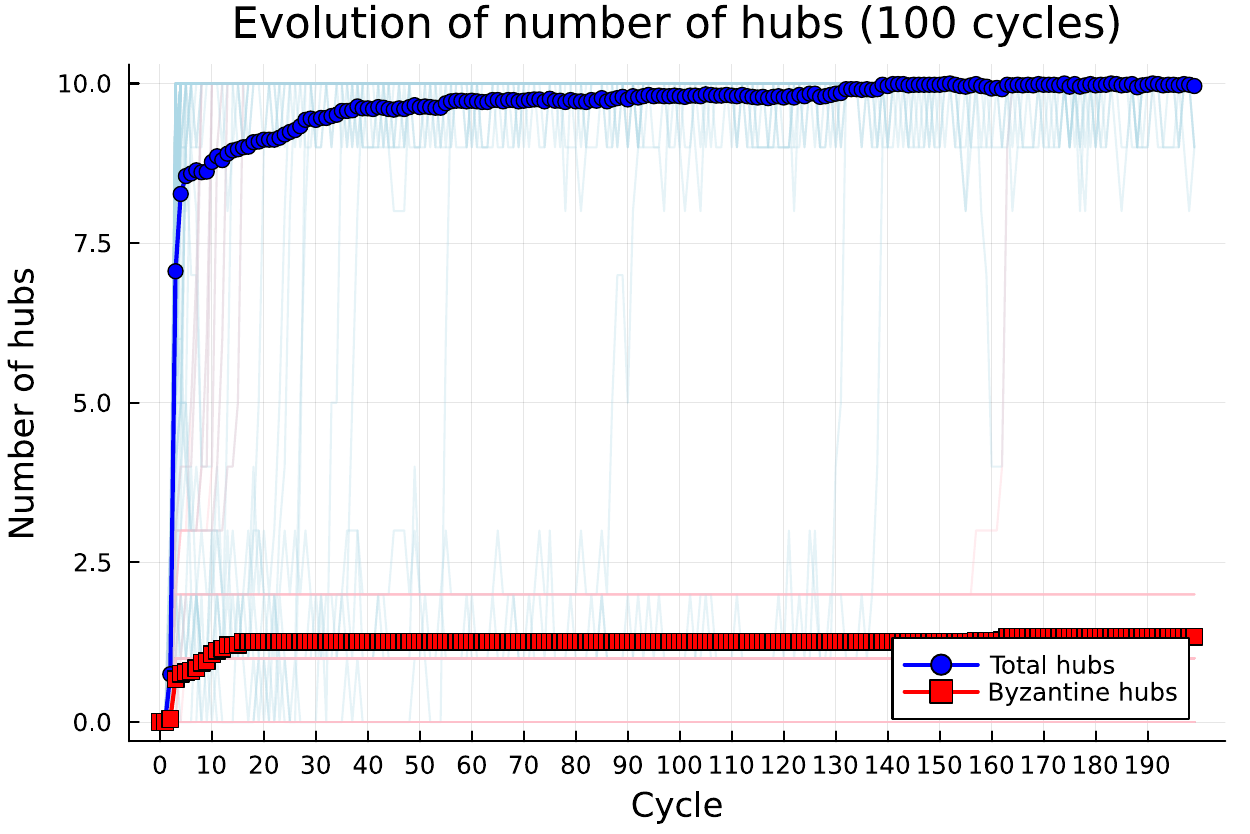}
\caption{Byzantine hub infiltration at 1\% participation rate showing byzantine infiltration but with honest hubs still being the majority.}
\label{fig:1percent_byzantine}
\end{figure}

\begin{figure}[htbp]
\centering
\includegraphics[width=0.45\textwidth]{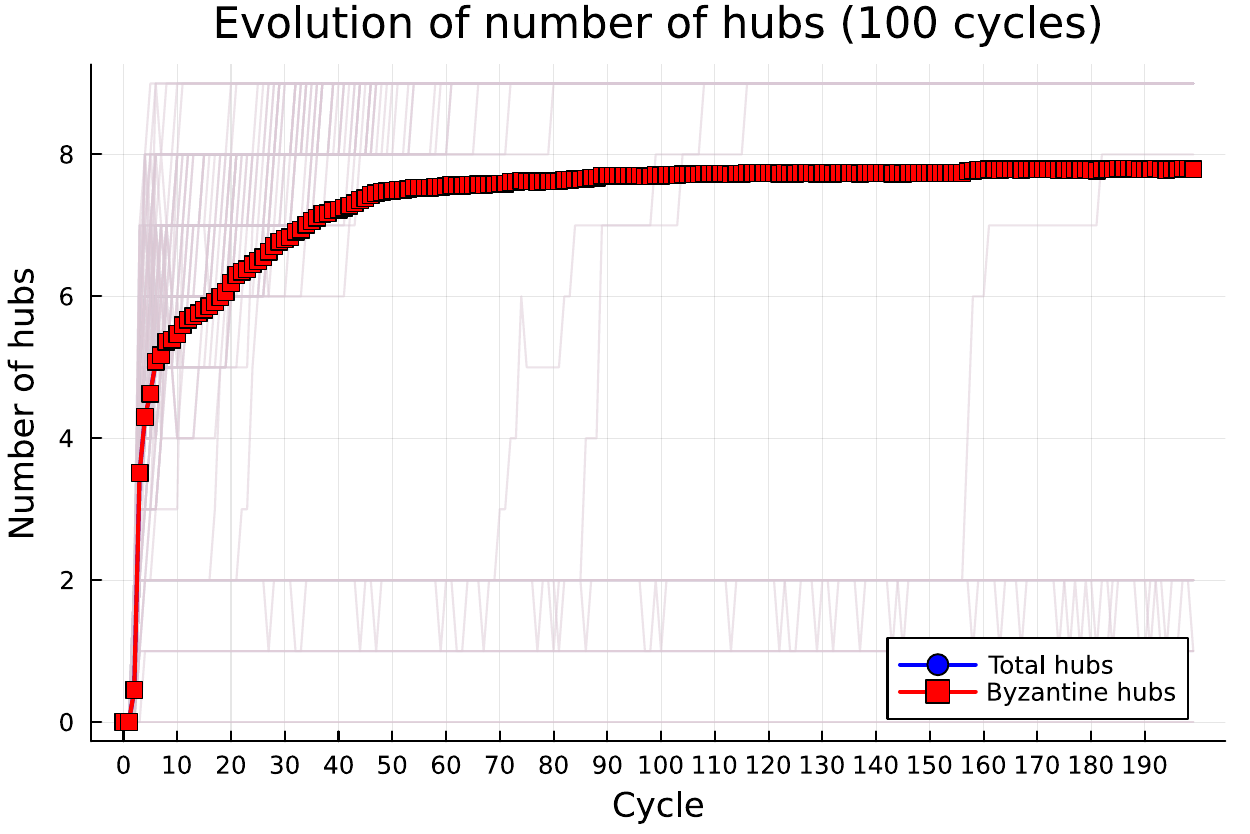}
\caption{Byzantine hub infiltration at 2\% participation rate demonstrating the critical vulnerability threshold where Byzantine nodes begin systematic hub capture.}
\label{fig:2percent_byzantine}
\end{figure}

\begin{figure}[htbp]
\centering
\includegraphics[width=0.45\textwidth]{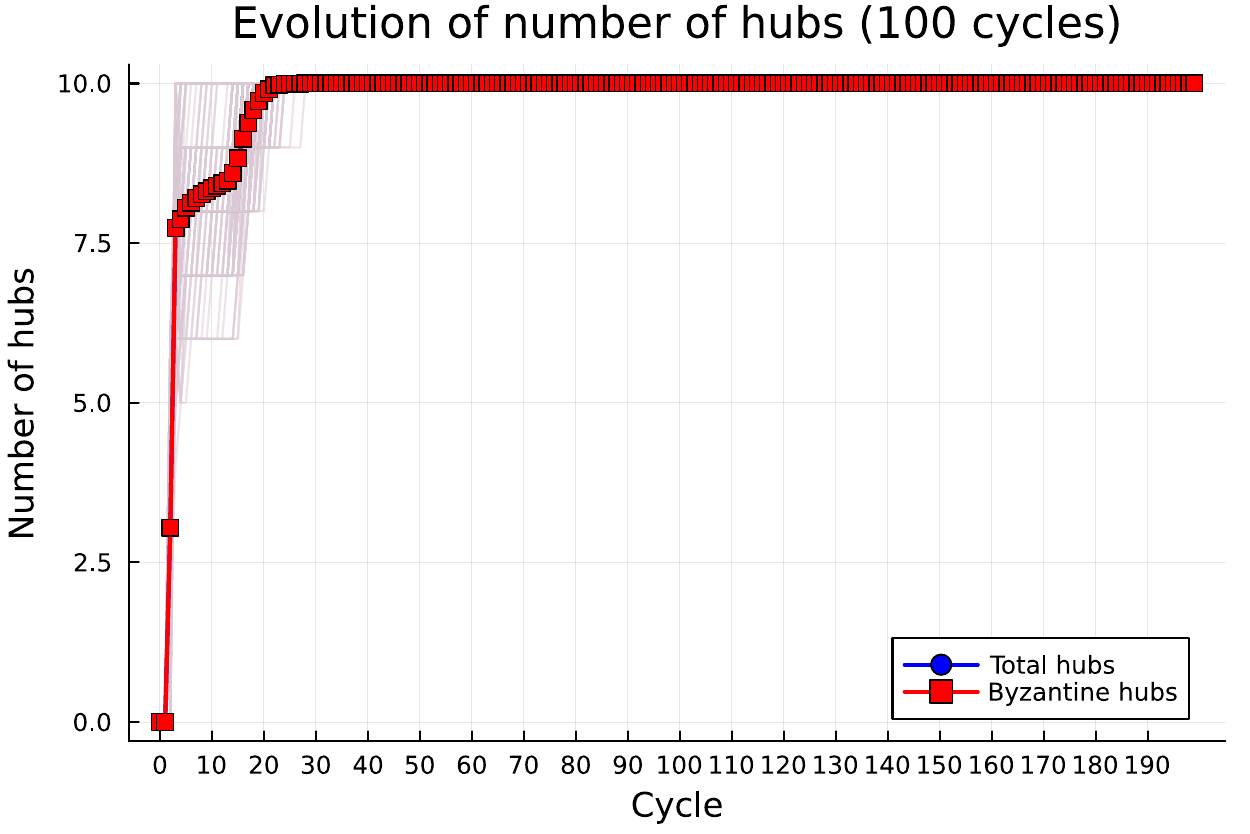}
\caption{Byzantine hub infiltration at 5\% participation rate where all the hubs are Byzantine nodes.}
\label{fig:5percent_byzantine}
\end{figure}

Figures~\ref{fig:1percent_byzantine},~\ref{fig:2percent_byzantine} and~\ref{fig:5percent_byzantine} illustrate the critical vulnerability threshold discovered in our experiments. 
A 1\% Byzantine rate, we have on average 1.34 hubs that are byzantines, meaning we went from 1\% byzantines nodes to 13.4\% byzantines hubs.
A sharp transition occurs around 2\% Byzantine participation, where Byzantine nodes begin to systematically dominate hub formation. This threshold appears to correlate strongly with the cache size parameter ($c = 20$), suggesting that when the number of Byzantine nodes approach or exceed the cache size, their coordinated responses can overwhelm the random sampling mechanisms. At 5\% Byzantine participation, all the 10 hubs are byzantines.

Our experimental evaluation reveals several critical findings regarding Elevator's Byzantine resilience:

The most significant discovery is the existence of a vulnerability threshold around 2\% Byzantine participation. 
Below this threshold, Elevator maintains strong resistance to hub infiltration, but beyond it, Byzantine nodes systematically capture hub positions. This threshold correlates with the cache size parameter (2\% of 1000 is 20, which is the size of the cache), suggesting that attackers need sufficient presence to influence the random sampling process effectively.

Independent Byzantine attacks prove largely ineffective, while coordinated attacks demonstrate high success rates. This finding indicates that Elevator's primary vulnerability lies in its susceptibility to coordinated misinformation rather than individual malicious behavior.

\subsection{Effectiveness of LIFT counter measure}

We evaluate our new protocol LIFT  effectiveness across different Byzantine participation rates using the same experimental setup as our vulnerability analysis.

\subsubsection{5\% Byzantine Participation} 

\begin{figure}[htbp]
\centering
\includegraphics[width=0.45\textwidth]{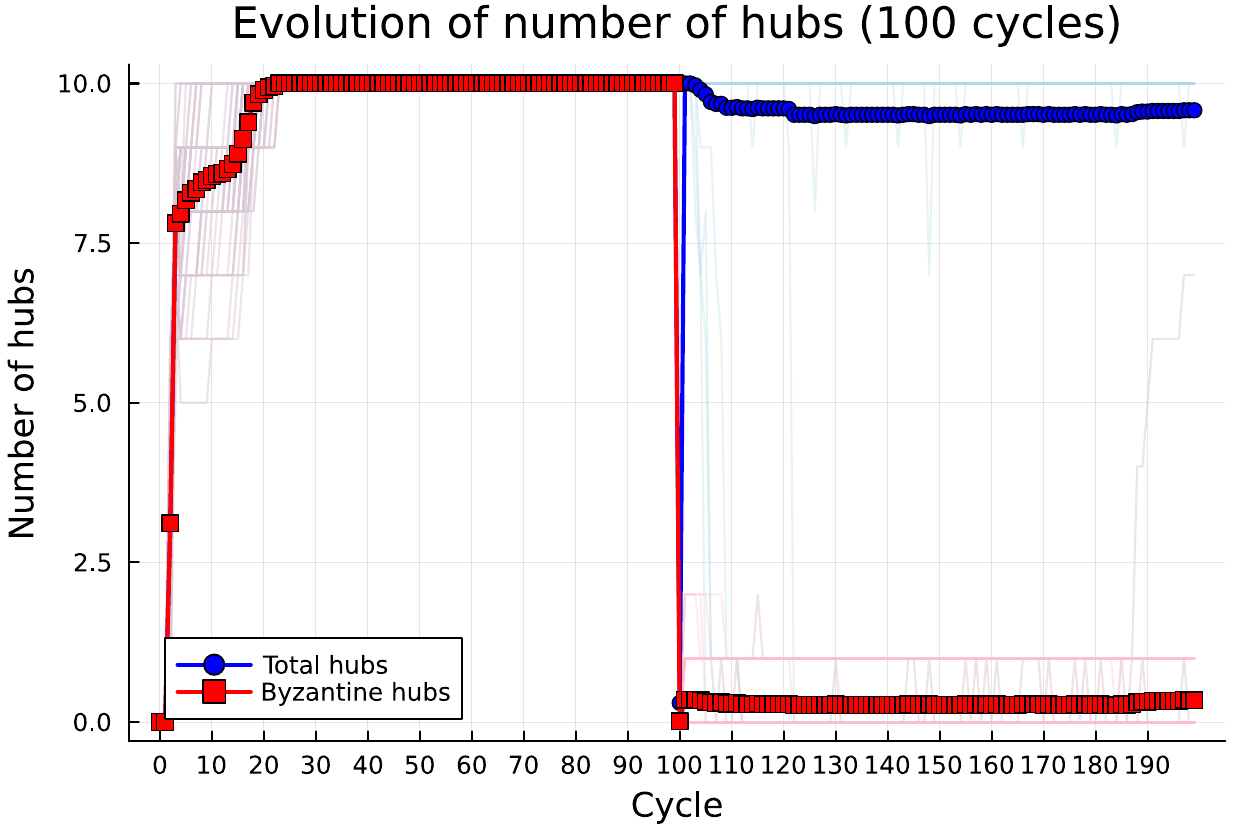}
\caption{Counter-attack effectiveness at 5\% Byzantine participation showing almost complete elimination of Byzantine hubs after cycle 100.}
\label{fig:counter_5percent}
\end{figure}

Figure~\ref{fig:counter_5percent} demonstrates complete effectiveness at this participation level. After counter-attack activation at cycle 100, Byzantine hubs are eliminated and remain at a low level throughout the remaining cycles, while the average total number of hubs decrease at 9.58. We have an average of 0.34 hubs that are Byzantine after the counter measure, so we went from 5\% Byzantine nodes to 3.4\% Byzantine hubs. This represents an almost complete recovery from Byzantine infiltration at the small cost of a slightly lower average number of hubs.

\subsubsection{10\% Byzantine Participation} 
\begin{figure}[htbp]
\centering
\includegraphics[width=0.45\textwidth]{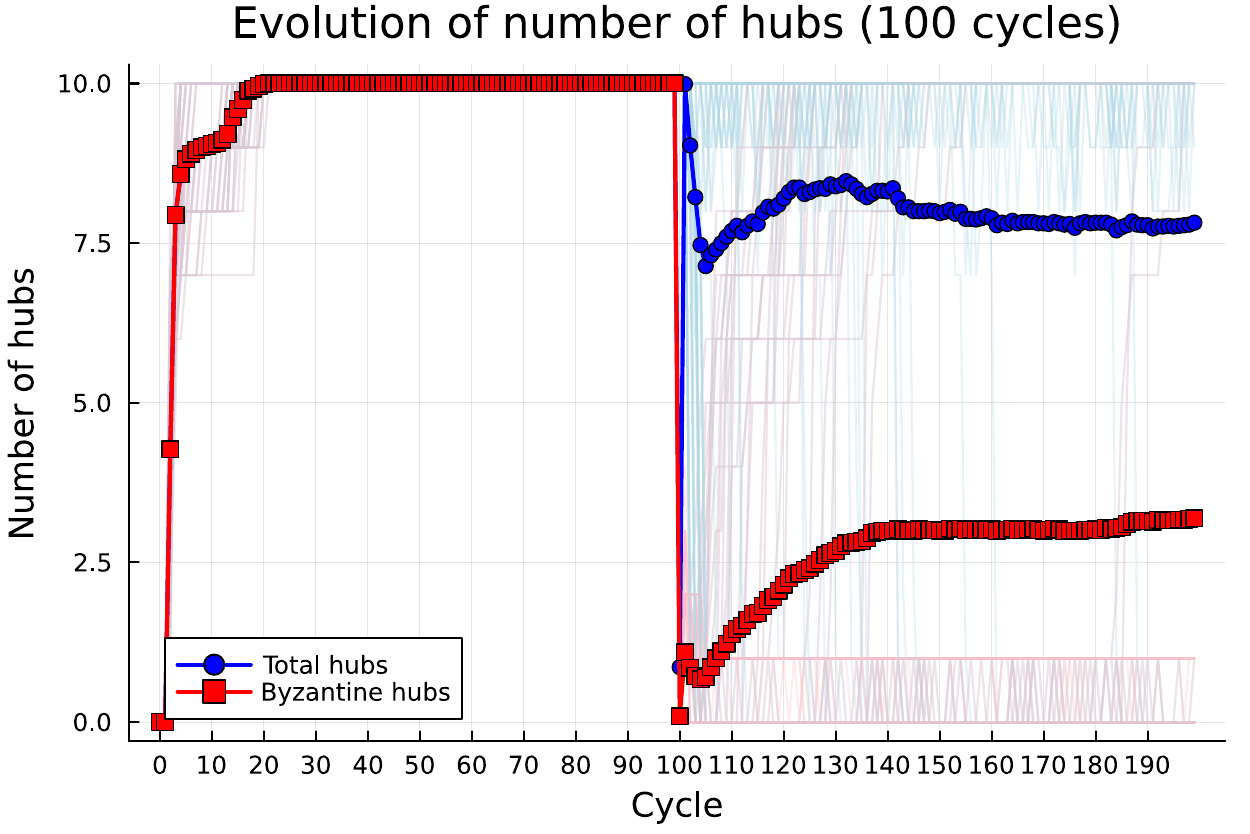}
\caption{Counter-attack effectiveness at 10\% Byzantine participation showing initial success followed by gradual Byzantine re-infiltration.}
\label{fig:counter_10percent}
\end{figure}

Figure~\ref{fig:counter_10percent} shows reduced effectiveness compared to the 5\% case. While the counter-attack initially eliminates Byzantine hubs at cycle 100, Byzantine nodes gradually reestablish themselves as hubs over subsequent cycles, reaching on average 3.19 Byzantine hubs by the end of the simulation. The total number of hubs is also decreasing, from 10 to 7.82 on average. This means that around 40\% of hubs are Byzantine, which is not entirely satisfactory, but with the majority of hubs being non-Byzantine, this remains acceptable.

\subsubsection{15\% Byzantine Participation} 
\begin{figure}[htbp]
\centering
\includegraphics[width=0.45\textwidth]{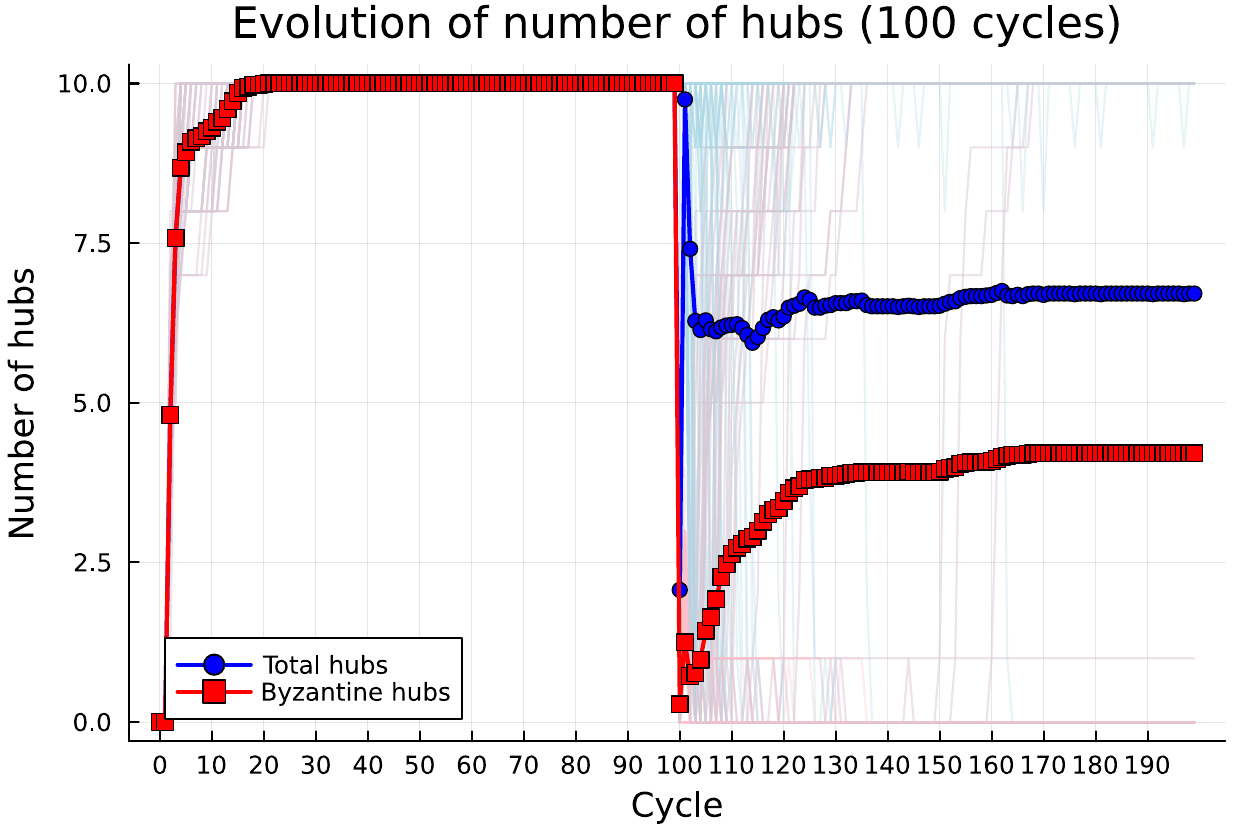}
\caption{Counter-attack effectiveness at 15\% Byzantine participation showing limited long-term effectiveness against high Byzantine participation.}
\label{fig:counter_15percent}
\end{figure}

Figure~\ref{fig:counter_15percent} shows that effectiveness decreases further at higher Byzantine percentages. While the counter-attack initially eliminates Byzantine hubs at cycle 100, Byzantine nodes gradually reestablish themselves as hubs in subsequent cycles, reaching on average 4.21 Byzantine hubs by the end of the simulation. The total number of hubs is also decreasing, from 10 to 6.71 on average. This means that approximately 62\% of hubs are Byzantine, indicating that our countermeasure is no longer effective.
\\

\subsubsection*{Summary}
The results demonstrate that the counter-attack successfully breaks Byzantine coordination at lower percentages (5\%) by introducing a deterministic selection process. However, effectiveness decreases as Byzantine participation increases—at 10\% and 15\% Byzantine participation, malicious nodes gradually re-infiltrate hub positions after the initial counter-attack activation. As the percentage of Byzantine nodes increases, the total number of hubs also decreases, indicating that some nodes are prevented by Byzantine nodes from remaining hubs.

An interesting and unexpected finding from the simulation results is that even after the countermeasure, the Byzantines continue to try to take over hubs and are partially successful. Their attacks also have the effect of preventing the full retention of all hubs in some cases. These results were not at all expected and therefore reduce the success of LIFT compared to the theoretical results expected to have an average of $\frac{B}{N}$ hubs that are Byzantine. These results lead us to conclude that we need to consider other more advanced methods to further reduce the influence of the Byzantines. Nevertheless, our method succeeds in significantly reducing the influence of Byzantines and has the advantage of not requiring a lot of resources, given that it is a one-shot solution. 
The proposed counter-attack addresses Elevator's primary vulnerability by introducing a deterministic redistribution mechanism based on immutable node identifiers. This requires synchronized activation at a predetermined cycle but avoids reliance on Byzantine-resistant communication.

While effective at reducing Byzantine influence, the mechanism assumes that the participants in the network do not change  and fixed activation point, making it sensitive to convergence timing and less effective under high Byzantine participation or network churn.

\section{Conclusions and Future Works}
\label{sec:conclusion}
This paper presents the first analysis of Byzantine attack resilience in the Elevator hub-sampling protocol. Our key finding is a critical vulnerability threshold at 2\% Byzantine participation—below this, Elevator resists attacks well, but beyond it, coordinated Byzantine nodes systematically capture all hub positions. This threshold correlates with the cache size parameter, suggesting attackers need sufficient presence to overwhelm the sampling process.

We showed that coordination is essential for successful attacks: independent Byzantine nodes have limited impact even at 10\% participation, while coordinated attackers succeed beyond the 2\% threshold. %

Our proposed new protocol (LIFT), with a counter-attack mechanism, successfully addresses these vulnerabilities by using deterministic hub redistribution based on immutable node identifiers. Results show impressive effectiveness at 5\% Byzantine participation and substantial improvement at higher rates.

Future improvements could include adaptive activation triggered by network behavior or Byzantine detection. Expanding the identifier space (e.g., $[0, N^k)$) could improve robustness, and retaining the initial hub list would allow consistent seed generation for newly joining nodes. Also, we considered specific attacks where the Byzantine nodes worked to infiltrate the hubs, studying other attacks (\emph{e.g.} DDoS, generating inconsistencies, etc.) would be an interesting path for future research. Our LIFT protocol assumes that node identifiers are assigned randomly, cannot be modified by Byzantine nodes, and are immutable. Removing some of those hypotheses is an interesting challenge.

Another area of research would be to evaluate the performance of our approach in terms of resilience to poisoning attacks, i.e., attacks where Byzantine nodes send false machine learning models. There has been work in the literature on poisoning attacks in decentralized networks~\cite{pham2024data}, and there are even peer sampling algorithms designed to be resilient to such attacks~\cite{belal2025granite}. It would be interesting to compare our LIFT protocol to these approaches.

\bibliographystyle{IEEEtran}
\bibliography{references}

\end{document}